\documentstyle[12pt,psfig]{article}

\begin{document}

{\footnotesize Functional Materials {\bf 5}, No. 3 (1998) 315--318.}

\begin{center}
{\bf EFFECTS OF NOISE AND NONLOCAL INTERACTIONS \\
IN NONLINEAR DYNAMICS OF MOLECULAR SYSTEMS} \\[4mm]

{\bf P.L. Christiansen$^1$, G.I. Gaididei$^2$, 
Yu.B. Gaididei$^3$, \\ M. Johansson$^4$,  
S.F. Mingaleev$^3$, and K.{\O}. Rasmussen$^5$} \\[4mm]

{\sl $^1$Department of Mathematical Modelling, Technical 
University of Denmark, \\ DK-2800 Lyngby, Denmark\\
$^2$Physics Department of  Kyiv University, 252022, Kyiv, Ukraine\\ 
$^3$Bogolyubov Institute for Theoretical Physics, 
252 143  Kyiv, Ukraine\\ 
$^4$Department of Physics and Measurement Technology, \\ 
Link{\"o}ping University, S-581 83  Link{\"o}ping, Sweden\\
$^5$Los Alamos National Laboratory, Los Alamos, New Mexico 87545,
USA}

\end{center}

\begin{center}
\parbox{110mm}{\small 
We show that the NLS systems with multiplicative noise, nonlinear 
damping and nonlocal dispersion exhibit a variety of interesting 
effects which may be useful for modelling the dynamical behavior of
one- and two-dimensional systems. 
}
\end{center}

\vspace{3mm}
{\bf 1 \ INTRODUCTION}
\vspace{3mm}

We consider the nonlinear Schr{\"o}dinger (NLS) system with 
cubic nonlinearity. 
This system models optical media, molecular thin films in the 
continuum limit, deep water waves and many other physical systems 
which exhibit weak nonlinearity and strong dispersion. The effects 
we shall discuss are observed in our earlier works 
\cite{magnus3,G-PRE96,gmcr} and cited therein. 

The motivation for studying the 2-D NLS equation with thermal 
fluctuations is our 
intension to model efficient energy transfer in Scheibe-aggregates. 
Our starting point is a two-dimensional Davydov model with nonlinear 
coupling between the exciton and phonon system, and white noise in 
the phonon system. In Section 2 we derive a single equation for the 
exciton system with 
multiplicative colored noise and a nonlinear damping term. In the 
continuum limit the collective coordinate approach indicates that 
an energy balance between energy input (from the noise term) and 
dissipation can be established. Thus, this model may describe the 
state of thermal equilibrium in the molecular aggregate. The coherent 
exciton moving on the aggregate \cite{moebius1,moebius2} is modelled 
by the ground state solution to the 2-D NLS equation, 
and the lifetime has been related to the collapse time of the ground 
state \cite{Peter}. For sufficiently strong nonlinearity the thermal 
fluctuations will slow down the collapse. As a result this lifetime 
increases with the variance of the fluctuations, i.e. the temperature.

In Section 3 the role of long-range dispersive interaction in 
the one-dimensional molecular system is investigated. 
Dispersive interactions of two types (the power and the 
exponential dependences of the interaction intensity
on the distance) are studied. If the intaraction decreases with 
the distance slowly, there is an interval of bistability where 
two stable stationary states: narrow, pinned states and broad, 
mobile states exist at each value of the excitation energy. 
For cubic nonlinearity the bistability of the solitons occurs 
already for dipole-dipole dispersive interaction. We demonstrate 
a possibility of the controlled switching between pinned and 
mobile states applying a spatially symmetric perturbation in the 
form of a parametric kick. The mechanism could be important for 
controlling energy storage and transport in  molecular systems.

\vspace{3mm}
{\bf 2 \ NOISE AND DAMPING}
\vspace{3mm}

Following the derivation given in Ref.\ \cite{magnus3}, we start by 
assuming that the coupled exciton-phonon system can be described by 
the following pair of equations 
\begin{equation}
i \hbar \dot \psi_n +\sum_{n'} J_{nn'}\psi_{n'} +\chi u_n\psi_n=0 \; ,
\label{mag1}
\end{equation}
\begin{equation}
M{\ddot u}_n +M\lambda {\dot u}_n +M\omega_o^2u_n-\chi|\psi_n|^2= 
\eta_n(t) \; .
\label{mag2}
\end{equation}
Here $\psi_n$ is the amplitude of the exciton wave function 
corresponding to site $n$ and $u_n$ represents the elastic degree of 
freedom at site $n$. Furthermore, $-J_{nn'}$ is the dipole-dipole 
interaction energy, $\chi$ is the exciton-phonon coupling constant, 
$M$ is the molecular mass, $\lambda$ is the damping coefficient, 
$\omega_0$ is the Einstein frequency of each oscillator, and 
$\eta_n(t)$ is an external force acting on the phonon system. To 
describe the interaction of the phonon system with a thermal 
reservoir at temperature $T$, $\eta_n(t)$ is assumed to be Gaussian 
white noise with zero mean and the autocorrelation function 
\begin{equation}
\langle \eta_n(t)\eta_{n'}(t')\rangle = 2 M \lambda k_B T 
\delta(t-t') \delta_{nn'} \; ,
\label{mag3}
\end{equation}
in accordance with the fluctuation-dissipation theorem ensuring 
thermal equilibrium.

In order to derive a single equation for the dynamics of the exciton 
system, we start by writing the solution to Eq.\ (\ref{mag2}) in 
the integral form. 
Neglecting all exponentially decaying transient terms and 
making the additional assumption that $\psi_n$ varies slowly in space 
and that only nearest-neighbor coupling $J$ is of importance, we 
obtain \cite{G-PRE96} in the continuum approximation for the continuous 
exciton field $\psi(x,y,t)\equiv \sqrt{V/J} e^{-4iJt/\hbar}\psi_n(t)$ the 
equation of motion 
\begin{eqnarray}
i \psi_t +\nabla^2\psi+|\psi|^2\psi-\Lambda\psi(|\psi|^2)_t+\sigma 
\psi=0 \; , 
\label{mag15}
\end{eqnarray}
where $x$ and $y$ are scaled on the distance $\ell$ between nearest 
neighbors, a noise density $\sigma(x,y,t)$ is not white, but strongly 
colored \cite{magnus3}, and time was transformed into the dimensionless 
variable: $Jt/\hbar \rightarrow t$. The nonlinear damping parameter 
$\Lambda=\frac{\lambda J}{\hbar \omega_0^2}$. 

It can easily be shown that in spite of the presence of the nonlinear 
damping and multiplicative noise terms in Eq.\ (\ref{mag15}), the 
norm, defined as 
\begin{eqnarray}
N=\int\int|\psi(x,y,t)|^2dxdy 
\label{mag17}
\end{eqnarray}
will still be a conserved quantity. 

To investigate the influence on the collapse process of the damping 
and noise terms in Eq.\ (\ref{mag15}), we will use the method of 
collective coordinates. To this end, we will make some simplifying 
assumptions. We will assume isotropy, which effectively reduces the 
problem to one space dimension with the radial coordinate 
$r=\sqrt{x^2+y^2}$. We also assume that the noise $\sigma$ can be 
approximated by radially isotropic Gaussian white noise. The validity of 
the approximation was discussed in Ref.\ \cite{magnus3}. Finally, 
we assume that the collapse process can be described in terms 
of collective coordinates using the following self-similar trial 
function for the exciton wave function $\psi(r,t)$  
\begin{eqnarray}
\psi(r,t)=A(t)\mbox{sech}\left(\frac{r}{B(t)}\right)e^{i\alpha 
(t) r^2} \; . 
\label{mag22}
\end{eqnarray}
This trial function, with three real time-dependent parameters 
$A$, $B$, and $\alpha$ determining the amplitude, width, resp.\ 
phase of the wave function, was used in Refs. \cite{magnus4,magnus6} 
to investigate the case when $\Lambda=0$ in Eq.\ (\ref{mag15}). 
The choice of this particular type of trial function can be motivated 
by regarding it as a generalization of the approximate ground state 
solution to the ordinary 2-D NLS found in Ref.\ \cite{magnus7}. From 
the definition (\ref{mag17}) of the norm, we immediately obtain the 
relation between amplitude and width 
$A(t) \sim \sqrt{N} / B(t)$. 
In analogy with the treatment for the undamped case in Ref.\ 
\cite{magnus6}, we find that it is possible to arrive to the 
following differential equation for the width $B$ of the exciton 
wave function  
\begin{eqnarray}
\ddot B=\frac{\Delta}{B^3}-\frac{\Gamma \dot B}{B^4}+ 
\frac{h(t)}{B^2} \; ,
\label{mag31}
\end{eqnarray}
where the constants $\Delta$ and $\Gamma$ are some functions of 
$N$ and $\Lambda$. Note that $\Delta$ and $\Gamma$ 
depend on the initial conditions via $N$, and that while $\Delta$ 
can be either positive or negative, $\Gamma$ is always positive. 
In the absence of noise and damping the collapse will occur if 
and only if $\Delta <0$. The white noise $h(t)$ has the 
autocorrelation 
\begin{eqnarray}
\langle h(t)h(t')\rangle = 2D\delta(t-t') \; , 
\label{mag32}
\end{eqnarray}
where $D$ is the dimensionless noise variance. 

Our numerical calculations \cite{G-PRE96} show that for 
$D<D_{\mbox{crit}}\simeq 0.15$, the effect of the noise is to delay 
the pseudo-collapse in terms of the ensemble average of the width, 
in analogy with the similar result obtained in Ref.\ \cite{magnus4} 
for the undamped case. For $D>D_{\mbox{crit}}$, we observe a 
non-monotonic behavior of $\langle B(t) \rangle$.  Initially, 
the average width will decrease in a similar way as when 
$D <D_{\mbox{crit}}$, but after some time $\langle B(t) \rangle$ 
will reach a minimum value and diverge as $t \rightarrow \infty$. 
This is due to the fact that for $D>D_{\mbox{crit}}$, the noise is 
strong enough to destroy the pseudo-collapse and cause dispersion 
for some of the systems in the ensemble. As $t \rightarrow \infty$ 
the dominating contribution to $\langle B(t) \rangle$ will come 
from the dispersing systems for which $B \rightarrow \infty$, and 
consequently  $\langle B(t) \rangle$ will diverge for 
$D >D_{\mbox{crit}}$. The minimum value of $\langle B(t) \rangle$ 
will increase towards $B(0)$ as $D$ increases.

\vspace{3mm}
{\bf 3 \ NONLOCAL INTERACTIONS} 
\vspace{3mm}

In the main part of the previous studies of the discrete NLS models 
the dispersive interaction was assumed to be short-ranged and a 
nearest-neighbor approximation was used. However, there exist physical 
situations that definitely can not be described in the framework of 
this approximation. The DNA molecule contains charged groups, with 
long-range Coulomb interaction ($1/r$) between them. The excitation 
transfer in molecular crystals and the vibron energy 
transport in biopolymers are due to transition 
dipole-dipole interaction with $1/r^3$ dependence on the distance, $r$. 
The nonlocal (long-range) dispersive interaction in these systems 
provides the existence of additional length-scale: the radius of the 
dispersive interaction. We will show that it leads to the bifurcative 
properties of the system due to both the competition between 
nonlinearity and dispersion, and the interplay of long-range 
interactions and lattice discreteness. 

In some approximation the equation of motion is the nonlocal 
discrete NLS equation of the form 
\begin{eqnarray}
i\frac{d}{d t}\psi_n+\sum_{m\neq n}J_{n-m}(\psi_m-\psi_n)+ 
|\psi_n|^2\psi_n=0 \; ,
\label{eq21}
\end{eqnarray} 
where the long-range dispersive coupling is taken to be either 
exponentially, $J_n=J\,e^{-\beta|n|}$, or algebraically, 
$J_n=J\,|n|^{-s}$, decreasing with the distance $n$ between 
lattice sites. The parameters $\beta$ and $s$ are introduced to cover 
different physical situations from the nearest-neighbor 
approximation ($\beta \rightarrow \infty, \; s \rightarrow 
\infty$) to the quadrupole-quadrupole ($s=5$) and dipole-dipole 
($s=3$) interactions. The equation (\ref{eq21}) conserves the number 
of excitations $N=\sum\limits_n |\psi_n|^2$. 

\begin{figure}[ht]
\begin{center}
\leavevmode
\psfig{figure=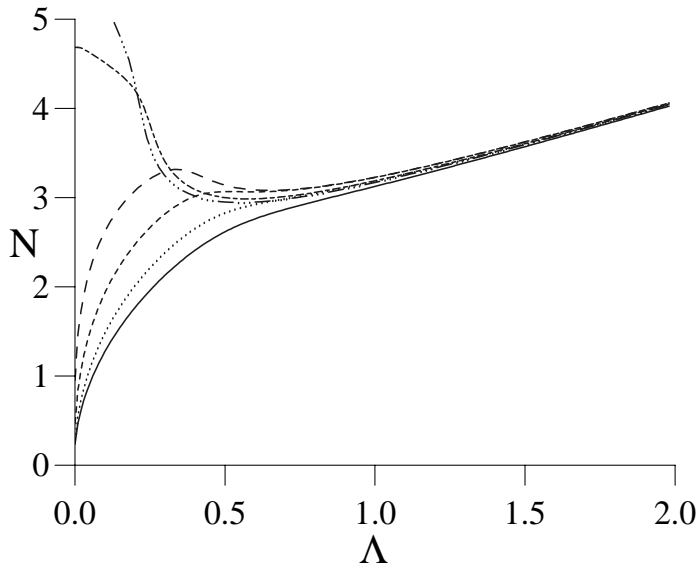,height=3.0in,angle=270}
\caption{Number of excitations, $N$, versus frequency, 
$\Lambda$, found numerically from Eq.\ (\ref{23}) for 
$s=\infty$ (full), 4 (dotted), 3 (short-dashed), 
2.5 (long-dashed), 2 (short-long-dashed),
1.9 (dashed-dotted).}
\label{fig2}
\end{center}
\end{figure} 

We are interested in stationary solutions of Eq.\ (\ref{eq21})
of the form $\psi_n(t)=\phi_n \exp (i \Lambda t)$ with a real shape 
function $\phi_n$ and a frequency $\Lambda$. This gives the 
governing equation for $\phi_n$ 
\begin{equation}
\label{23}
\Lambda \phi_n=  \sum_{m \neq n} J_{n-m} (\phi_m-\phi_n)+ 
\phi_n^{3} \; .
\end{equation}
Figure \ref{fig2} shows the dependence $N(\Lambda)$ obtained from 
direct numerical solution of Eq.\ (\ref{23}) for algebraically 
decaying $J_{n-m}$. A monotonic function is obtained only for 
$s>s_{cr}$. For $2<s<s_{cr}$ the dependence becomes nonmonotonic (of 
${\cal N}$-type) with a local maximum and a local minimum. These 
extrema coalesce at $s=s_{cr} \simeq 3.03$. For $s<2$ the local 
maximum disappears. The dependence $N(\Lambda)$ obtained analytically 
using the variational approach is in a good qualitative agreement with 
the dependence obtained numerically (see \cite{gmcr}). Thus the main 
features of all discrete NLS models with dispersive interaction 
$J_{n-m}$ decreasing faster than $|n-m|^{-s_{cr}}$ coincide 
qualitatively with the features obtained in the nearest-neighbor 
approximation where only one stationary state exists for any number 
of excitations, $N$. However in the case of long-range nonlocal NLS 
equation (\ref{eq21}), i.e.\ for $2<s<s_{cr}$, there exist for each 
$N$ in the interval $[N_{l}(s), N_{u}(s)]$ three stationary states 
with frequencies $\Lambda_{1}(N) < \Lambda_{2}(N) < \Lambda_{3}(N)$. 
In particular, this means that in the case of dipole-dipole 
interaction ($s=3$) multiple solutions exist. It is noteworthy that 
similar results are also obtained for the dispersive interaction of 
the exponentially decaying form. In this case the bistability takes 
place for $\beta\,\leq\,1.67$.  According to the theorem which was 
proven in \cite{lst94}, the necessary and sufficient stability 
criterion for the stationary states is $dN /d\Lambda > 0$. Therefore, 
we can conclude that in the interval $[N_{l}(s), N_{u}(s)]$ there 
are only two linearly stable stationary states: $\Lambda_{1}(N)$ 
and $\Lambda_{3}(N)$. The intermediate state is unstable since 
$dN /d\Lambda < 0$ at $\Lambda=\Lambda_2$. 

The low frequency states 
are wide and continuum-like while the high frequency solutions 
represents intrinsically localized states with a width of a few 
lattice spacings. It can be shown 
that the existence of two so different soliton states for one value 
of the excitation number, $N$, is due to the presence of two different 
length scales in the system: the usual scale of the NLS model which 
is related to the competition between nonlinearity and dispersion 
(expressed in terms of the ratio $N/J$ ) and the range of the 
dispersive interaction $\xi$. 

Having established the existence of bistable stationary states in 
the nonlocal discrete NLS system, a natural question that arises 
concerns the role of these states in the full dynamics of the model. 
In particular, it is of interest to investigate the possibility of 
switching between the stable states under the influence of external 
perturbations, and to clear up what type of perturbations can 
be used to control the switching. Switching of this type is important 
for example in the description of nonlinear transport and storage of 
energy in biomolecules like the DNA, since a mobile continuum-like 
excitation can provide action at distance while the switching to a 
discrete, pinned state can facilitate the structural changes of the 
DNA \cite{geor96}. As it was shown recently in \cite{mj98}, switching 
will occur if the system is perturbed in a way so that an internal, 
spatially localized and symmetrical mode ('breathing mode') of the 
stationary state is excited above a threshold value.



\end{document}